\documentclass[journal]{IEEEtran}
\ifCLASSINFOpdf
\else
\fi
\usepackage{amsmath}
\usepackage{amssymb}
\usepackage{cite}
\usepackage{url}
\usepackage{float}
\usepackage{subfig}
\usepackage{graphicx}
\usepackage{subfig}
\usepackage{footnote}
\newcommand{\cov}{\textrm{Cov}}

\def\thick#1{\hbox{\rlap{$#1$}\kern0.25pt\rlap{$#1$}\kern0.25pt$#1$}}

\def\by{{\bf y}}

\begin{document}
%
\title{A New Approach to Customization of Collision Warning Systems to Individual Drivers}
%
%
%

\author{Ali~Rakhshan,~\IEEEmembership{Student Member,~IEEE,}
        ~Evan~Ray
        and Hossein~Pishro-Nik,~\IEEEmembership{Member,~IEEE,}.
\thanks{A. Rakhshan and H. Pishro-Nik are with the Department
of Electrical and Computer Engineering, University of Massachusetts, Amherst, (e-mails: {arakhshan,
pishro}@ecs.umass.edu)
MA, 01002 USA.}
\thanks{E. Ray is with the Department of Mathematics and Statistics, University of Massachusetts, Amherst, (e-mail: ray@math.umass.edu).}
\thanks{
}}

%
%

\markboth{IEEE TRANSACTIONS ON INTELLIGENT TRANSPORTATION SYSTEMS}%
{Shell \MakeLowercase{\textit{et al.}}: Bare Demo of IEEEtran.cls for Journals}
%



\maketitle

\begin{abstract}
This paper discusses the need for individualizing safety systems and proposes an approach including the Real-Time estimation of the distribution of brake response times for an individual driver. While maintaining high level of safety, the collision warning system should send ``tailored'' responses to the driver. This method could be the first step to show that safety applications would potentially benefit from customizing to individual drivers' characteristics using VANET. Our simulation results show that, as one of the imminent and preliminary outcomes of the new improved system, the number of false alarms will be reduced by more than $40\%$. We think this tactic can reach to even beyond the safety applications for designing the future innovative systems.
\end{abstract}

\begin{IEEEkeywords}
Intelligent Transportation Systems, VANET, Perception-Reaction (P-R) time, Collision avoidance, Safety.
\end{IEEEkeywords}

%
\IEEEpeerreviewmaketitle

\section{Introduction\protect\footnote{This work was supported by the NSF under CCF 0844725.}}

\begin{figure}[!t]
	\centering
\includegraphics[width=2.5in]{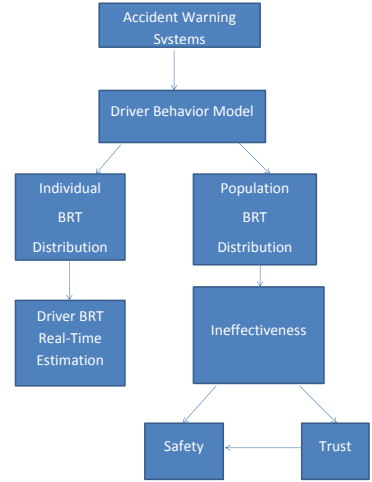}
		
	\caption{General Scheme: Collision Warning Systems}
	\label{fig:system}
\end{figure}

Despite the increases in safety introduced into the automobile, at latest count (2010) the number
of deaths is over 30,000, the number of injuries is over two million, and the number of crashes is
over ten million \cite{USDoT:TrafficFatalities}. Some of these accidents could have been prevented or reduced in severity if the drivers involved had been warned in time to slow down or steer away to avoid the accident. To address this problem, collision warning systems hold great promise, Fig. \ref{fig:system}. 
The true potential of the various classes
of warning systems to reduce crashes is seriously compromised by three interrelated factors:
\begin{itemize}
\item
The algorithms used to trigger a warning are largely ineffective when they are not adapted
to the individual driver and vehicles involved directly in a crash.
\item
 Warning algorithms
have relied for the most part on the behavior of threat vehicles immediately ahead and to the side.
\item The driver often fails to trust the warning even when it is issued in time to avoid
a crash.
\end{itemize}
Radical improvement in the effectiveness of collision warning systems are now possible due to
the progress that is being made in vehicular ad hoc networks (VANET). Vehicular ad hoc networks potentially allow all vehicles to communicate with each other (V2V or vehicle to vehicle communication) and with technologies embedded in the infrastructure that transmit crash relevant information (V2I or vehicle to infrastructure communication).

The effectiveness of warnings depends on how much time the driver needs to react. Therefore, to be as effective as possible, accident warning systems should be tailored to the specific characteristics of the driver.  He or she could be vigilant or distracted; could perceive and react soon to an event or might have a longer perception-reaction time; could be aggressive in acceleration/deceleration or could be smoother in those. An important aspect of  the specific characteristics of the driver is the distribution of brake response times (BRT) for each particular driver.  The BRT is the time elapsed between a stimulus such as a lead car braking or traffic signal changing color and a braking response by the driver. Since existing collision warning algorithms don't use the BRT distribution of individuals, drivers with different BRT in the same scenario receive the same warnings. Clearly, this approach isn't optimal for design of safety systems. The most important contributions of this paper are:
\begin{enumerate}
\item
Proposing a method for Real-Time estimation of the distribution of brake response times for an individual driver using data from a VANET system which has information about the positions, velocities, and accelerations of cars on the roads, road configurations, and the status and position of traffic signals.
\item
Using the estimated distribution to customize warning algorithms to an individual driver's characteristics.
\end{enumerate}

The paper is organized as follows.  In section \ref{section:Related Work} we review the relevant literature formally defining the BRT and related quantities, discussing factors that affect drivers' BRTs, and outlining several methods that have been proposed to estimate a driver's BRT.
  Section \ref{rtest} and \ref{estimation} outline methods that can be used to estimate BRTs and what the distribution of a driver's BRTs would be if he or she did not intentionally delay braking, respectively. Our concluding remarks are discussed in section \ref{conc}.

\section{Related Work}\label{section:Related Work}
\subsection{Basic Ideas: Perception-Reaction Times and Brake Response Times}

\begin{figure}[!t]
	\centering
\framebox{\parbox{3in}{\includegraphics[width=3in]{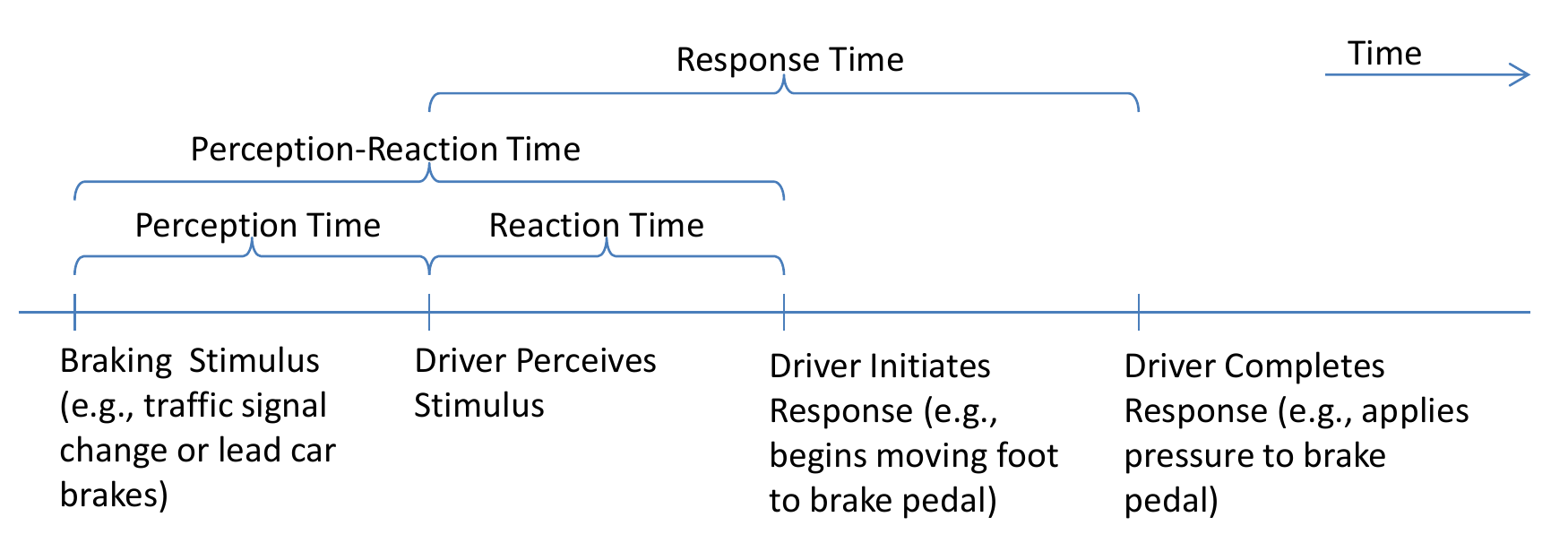}
}}
		
	\caption{The scheme for defining perception reaction times as given by Koppa \cite{Koppa:HumanFactors}.}
	\label{fig:KoppaPRTIllustration}
\end{figure}

The time required to respond to a stimulus can be divided into several distinct phases.  One such division is given by Koppa \cite{Koppa:HumanFactors}.  He defines the perception time as the amount of time it takes for an individual to recognize that an event has occurred.  The reaction time is then the time elapsed from detection of a stimulus to the start of a response.  The response time includes the reaction time as well as the time required to complete the response.  The perception-reaction time or brake reaction time is the time required to perceive and initiate a reaction to the stimulus.  These divisions are illustrated in Fig. \ref{fig:KoppaPRTIllustration}.

\begin{figure}[!t]
\centering
\framebox{\parbox{2.5in}{\includegraphics[width=2.5in]{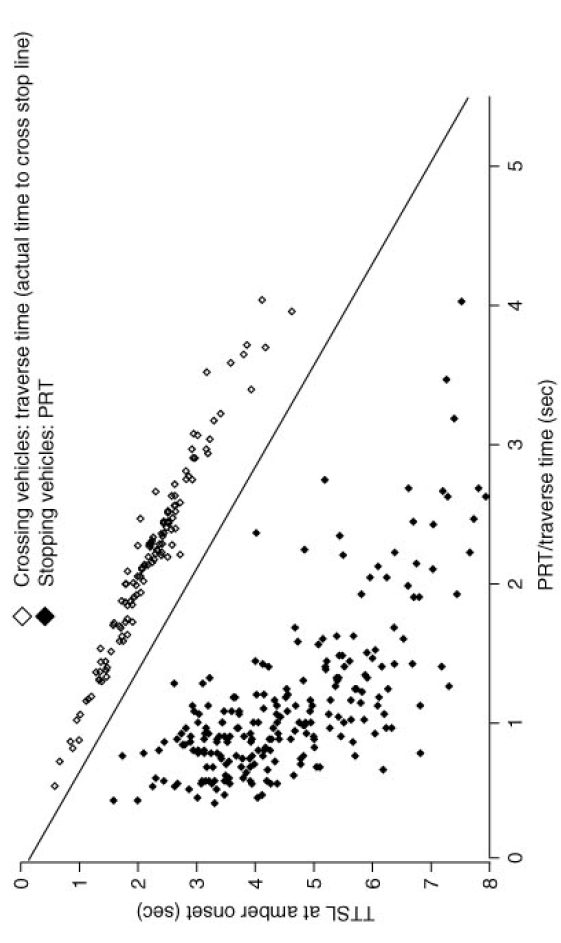}
}}
\caption{Rotated plot from Goh and Wong of observed brake reaction times (PRT in their terminology) vs. time headway to traffic signal \cite{GohWong:DriverPRTDuringSignalChange}.  Points above the diagonal line correspond to cars that did not stop at the intersection.}
\label{fig:GohWongRTPlot}
\end{figure}

There is some ambiguity in this definition of the reaction and response times in that we must specify what is meant by the response.  Commonly in driving studies, the response is operationally defined to be the act of braking \cite{Green:HowLongToStop}.  This operational definition is convenient because it is relatively easy to measure when the brakes have been applied.  However, a difficulty with this definition is that a driver may intentionally delay braking, for instance if there is a large space between the driver and a traffic signal or leading car.  This means that measured response times may be larger than the drivers' ``true'' response times \cite{GohWong:DriverPRTDuringSignalChange}, \cite{Rakha:DriBeh-conf}. This delay is illustrated in the data plot reproduced in Fig. \ref{fig:GohWongRTPlot}, which is taken from an article by Goh and Wong \cite{GohWong:DriverPRTDuringSignalChange}.  We have rotated the plot to clarify that we view time headway as the independent variable and response time as the dependent variable.  In this plot, the horizontal axis shows the driver's time headway to a traffic signal at the time it turned from green to yellow and the vertical axis shows the measured brake reaction time for drivers who braked (or the actual time to pass the signal for those drivers who ran the light).  We see that when the driver is a larger distance from the traffic signal, their measured brake response time is larger -- likely because they chose to delay braking.

\begin{figure}[!t]
	\centering
\framebox{\parbox{3in}{\includegraphics[width=3in]{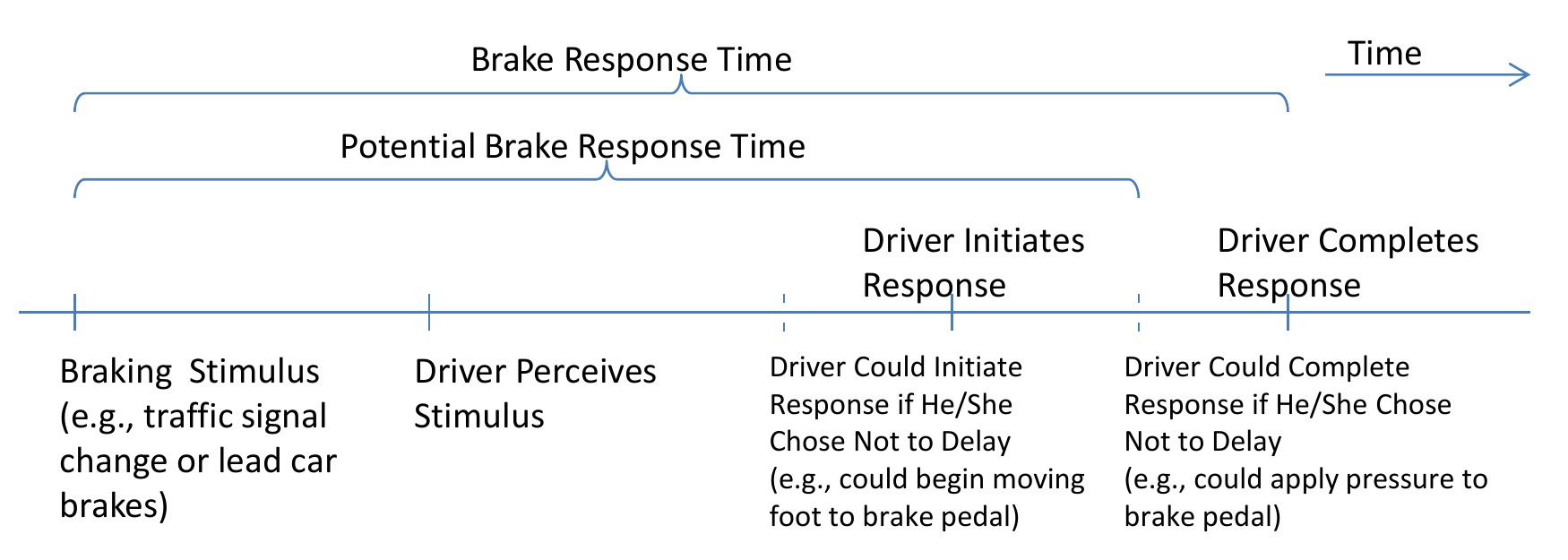}
}}
	\caption{An illustration of the potential brake response time and brake response time.}
	\label{fig:OurPRTIllustration}
\end{figure}

In this paper we define the potential brake response time (PBRT) as the time that a driver could have braked in if he or she did not choose to delay braking, which is the relevant quantity for the purposes of an accident warning system.  We will use the term ``brake response time'' (BRT) to refer to the observed quantity, the time elapsed between a stimulus such as a traffic signal color change and when the driver applies pressure to the brake pedal.  These definitions are illustrated in Fig. \ref{fig:OurPRTIllustration}.  The estimation of BRT and PBRT both present technical difficulties.  We review methods that have been proposed to estimate these quantities by previous researchers in the next two subsections.

Virtually every study to examine reaction times has found that the population distribution of reaction times is skewed right and several have shown that it is well approximated by a lognormal distribution 
\cite{Koppa:HumanFactors}, \cite{Green:HowLongToStop}, \cite{GohWong:DriverPRTDuringSignalChange}, \cite{MaxwellWood:ReviewTrafficSignal}, \cite{WortmanMatthias:EvalDriverBehavior}, \cite{ZhangBham:EstDriverRT}, \cite{Rakha:DriBeh}. A close examination of the plot in Fig. \ref{fig:GohWongRTPlot} above indicates that the distribution of BRTs is also skewed right at a fixed value of time headway.  We will make use of this observation later in our data analysis.

\subsection{Estimation of Brake Response Time from Car-Following Data}
\label{litrev:BRTEst}

Several previous studies have examined how BRT can be estimated from car-following data automatically.  Here we review several of these methods, focusing in particular on the effectiveness of these algorithms for obtaining an accurate estimate of the BRTs to several distinct events and on the feasibility of implementing them with the limited computational resources available in an on-board computer system in a car.

The main ideas we build on in this paper were proposed by Zhang and Bham \cite{ZhangBham:EstDriverRT}.  Their method is based on intuitive reasoning about the relationships between the distances, speeds, and accelerations of two cars when the following car reacts to an action taken by the lead car.  The starting point in their algorithm is to identify two cars which go for a period of at least 4 seconds in which they are separated by less than or equal to 250 feet and their speeds are within 5 ft/s, or 1.52 m/s.  These cars are said to be in a \textit{steady state}.  They then observe a time A when the the distance between the cars decreases or increases while the follower has an acceleration rate of $\leq 0.5 \text{ft/s}^2$.  This change in distance between the cars is caused by acceleration or deceleration of the leader.  Next they find the time B when the follower decelerates or accelerates at a rate $> 0.5 \text{ft/s}^2$.  The difference between times A and B is then an estimate of the follower's BRT.
The advantages of this method are that it is intuitively reasonable, relatively easy to implement, and it yields reasonable reaction time estimates.  However, the requirement that the cars be in steady state is restrictive.
To obtain more information about drivers' reaction times, it would be helpful to extend this approach to estimate reaction times in other situations than the steady state.

Another method for BRT estimation was proposed by Ma and Andr\'easson and is based on techniques designed to find the lag between two linearly related time series \cite{MaAndreasson:ReactionDelayEst}.  The inputs to this technique are two time series $\{x_t\}$ and $\{y_t\}$-- one which represents a stimulus, such as the difference in speed between the two cars, and another which captures the response of the follower, such as that car's acceleration.  The basic idea of the method is to examine the covariance between the time series in the frequency domain, as measured by the coherency.
There are two major limitations to this approach that make it inappropriate for our purposes.  First, the coherency measures linear association between the two time series.  If acceleration is not a linear function of the difference in speeds between the cars, this approach may not measure the lag between the time series appropriately.  Second, this method does not allow us to estimate separate BRTs to separate events in a natural way.  Instead it views the entire output time series as being a linear function of the entire input time series and estimates an overall time delay.  To estimate separate BRTs in order to build up a distribution of response times, we would have to manually divide the time series into shorter pieces corresponding to each reaction event.

A third approach was taken by Ahmed, who specified a reaction time distribution as part of a larger model of car-following behavior, and estimated all parameters of this model jointly through maximum likelihood techniques \cite{Ahmed:ModelingDrivers}.
However, the maximum likelihood estimates had to be obtained numerically, which is computationally intensive due to the complexity of the model.  Therefore, this method would not be practical to implement in an accident warning system where the BRT distribution must be obtained with limited computing resources. Furthermore, one of the desired requirements for the warning systems is to use the individual perception reaction time data online. In other words, the model needs to become more accurate as more information becomes available from VANET system. However, based on most of the current methods we cannot update the algorithm in Real-Time.

\subsection{Estimation of Potential Brake Response Times from Observed Brake Response Times}
\label{litrev:PRTEst}

Three previous studies have addressed the problem of estimating the distribution of ``true'' reaction times based on observed brake response times.  All of these studies examined this problem in the context of traffic signals, and focused on estimation of population distributions, rather than distributions of response times for a particular individual.

Maxwell and Wood simply used the mode response time as a point estimate for the average brake response time in a population, arguing that this measure would be less sensitive to large reaction times that include a delay \cite{MaxwellWood:ReviewTrafficSignal}.  Goh and Wong take a more sophisticated approach \cite{GohWong:DriverPRTDuringSignalChange}.  They define a transitional zone (TZ) based on the time headway between the driver and the traffic signal at the time that it changes to yellow.  This TZ is ``an empirically calibrated range of time headways suitable for identifying drivers with realistic stop-or-cross decisions" \cite{GohWong:DriverPRTDuringSignalChange}.  Essentially, to estimate response times they limit the sample to those cars with a time headway of $\leq$ 4 seconds.  Nearly all cars that chose not to stop at the light were within the 4-second threshold; thus, this threshold includes cars with a ``real'' choice between stopping and continuing on.  This is illustrated in Fig. \ref{fig:GohWongRTPlot} above.

This analysis does suffer from some limitations.  First, by restricting the sample to those cars within the TZ, they lose the information contained in those other data points.  This is a particularly critical problem in our application, where we wish to learn about response times for a particular driver.  We may not have the chance to observe response times very frequently for a single driver; it would therefore be helpful to be able to use all observed data points rather than just those with a time headway of 4 seconds or less.  Second, although the relationship between time headway and BRT is reduced when the sample is restricted to cars with time headway of $\leq$ 4 seconds, a relationship can still be seen in the plot in Fig. \ref{fig:GohWongRTPlot}.  This suggests that some of the measured response times may still include a delay even within the TZ.
In this paper, we pursue the idea that there is a significant relationship between BRT and these other variables, since multiple studies have supported this claim.

\section{Proposed Method for Brake Response Time Estimation}
\label{rtest}

For the purposes of the accident warning system, we wish to learn about the distribution of response times for an individual when the car in front of them brakes.  As discussed in section \ref{litrev:BRTEst}, Zhang and Bham have proposed an effective method for estimating BRT when the cars are in steady state.  However, this situation may be relatively rare in real-life driving situations, so that we may not make many observations of the BRT for an individual driver under this setting.  Therefore, in practice it could be difficult to learn about the distribution of response times using only the method proposed by Zhang and Bham.  Our proposed approach is to establish relationships between the distributions of reaction times under different circumstances.  This will allow us to use measures of a driver's response times under a variety of circumstances to estimate the distribution of an individual's response times when the car in front of them brakes.

We will attempt to measure brake response times in three settings:
\begin{enumerate}
	\item The cars are in steady state and the leader brakes.
	\item The cars are not in steady state, the follower is driving faster than the leader, and the leader brakes.
	\item The car approaches a traffic signal which changes from green to yellow.
\end{enumerate}

In this section we discuss specific ideas for reaction time estimation in each of these settings
 For now we concentrate on methods to obtain a point estimate for a driver's BRT to a particular event.  Methods to combine these point estimates to estimate the distribution of PBRTs will be discussed in the next section.

Although it is not mentioned in any of the algorithms below, we suggest that response times should only be recorded if the driver is travelling faster than some cutoff speed such as 20 miles per hour.

\subsection{Steady State, Leader Brakes}
In this case we use the algorithm developed by Zhang and Bham:

\begin{enumerate}
	\item Identify when a pair of cars is in steady state for 4 seconds:
		\begin{itemize}
			\item Separated by $\leq 250$ ft.
			\item Speed and acceleration of leader and follower are equalized (speeds must be within $\pm 5\text{ft/s} = 1.52\text{m/s}$).
			\item In the text, it is not clear whether specific limits are placed on acceleration, but it seems clear that if the distance and speed conditions are satisfied for 4 or more seconds, the cars' accelerations must be approximately equal.  However, there does seem to be a limit on the follower's acceleration of $0.5 \text{ft/s}^2$, from the second step below.
		\end{itemize}
	\item Observe a time A when the the distance between the cars starts to decrease while the follower has an acceleration rate of $\leq 0.5 \text{ft/s}^2$.  This change in distance between the cars is caused by acceleration or deceleration of the leader.
	\item Observe the time B when the follower decelerates at a rate $> 0.5 \text{ft/s}^2$.
\end{enumerate}

Zhang and Bham do not specify how they determined when the distance between the cars had started to decrease for step 2 in this algorithm.  Several methods are possible.  One simple idea is to determine at each time point whether the distance between the cars is less than it was at the previous measurement.  If this is sustained for a sufficient length of time (such as a quarter-second), the starting point A is the time at which the distance first started decreasing.  
If limitations of the measurement instrumentation mean that we may observe an increase or no change in the distances between the cars for one time point when they are actually decreasing, this approach could be replaced by regressing distance on time over a quarter-second period to determine if they have a negative association on average over that time.

\subsection{Not In Steady State, Leader Brakes}

In theory, it seems likely that a similar technique to the above can be used when the drivers are not yet in steady state and the lead car brakes.  Note that we might only expect to observe a response in this situation if the follower is travelling at a higher speed than the leader.  Also, the follower and leader should be near enough to each other that the follower will need to respond to the leader's braking action.  For example, we could measure response times only if the time headway between the leader and the follower is less than 10 seconds at the time that the leader brakes.

The key problems are selecting what measures to use in determining that the leader has braked and that the follower has responded.  For deciding whether the leader has braked, it may be easiest to make use of the vector of accelerations of the lead car, and use a threshold value to decide when the leader has braked.  We could simply use the value $-0.5 ft/s^2$ which was used above to detect when the following car reacted in the steady state setting.

To determine when the follower has responded, we would recommend first finding when a response has occurred in driving simulation trials by manually looking at the speed and acceleration profiles.  This should allow you to select what variables to use to measure the response.  One possibility that seems reasonable is a reduction in the acceleration of the follower.

Once this or some other similar quantity is determined to be the appropriate variable to use to detect the follower's response, we will again need to choose what cutoff value for that variable indicates that the response has occurred.  For example, we would need the cutoff value $c$ such that when the reduction in acceleration is less than $c$, we say that the follower has responded.  To find the value $c$ we could do a grid search, choosing $N$ candidate values $c_1, \ldots, c_N$ and running the classification code for each value $c_i$.  For values of $c_i$ which are too close to 0, the threshold will be exceeded easily and the algorithm will say the response time was shorter than the manually determined value.  For values of $c_i$ which are too far from 0, the threshold will be exceeded infrequently, and some of the manually determined responses will be missed.  The objective is to select a value $c_i$ such that the results of the classification algorithm best match the manual classification results.  This could be done informally, or formally by choosing $c$ to minimize a function such as the sum of squared differences between the manually determined response time and the algorithmically determined response time.

\subsection{Traffic Signal Changes from Green to Yellow}

There are several factors to consider when estimating a driver's brake response time to a traffic signal change.  First, we should only expect the driver to respond to the signal change if they are within a reasonable distance of the signal.  For that reason, we suggest a cutoff of 10 seconds in the time headway from the driver to the signal at the time it changes.  Second, we should not record a response time if there is an intervening car between the driver and the traffic signal that also responds to the signal change.  Finally, we should not record a response time if the driver turns at the intersection with the signal.  This would not be an accurate measure of the driver's response time since they would likely have been prepared to stop anyways.

We propose the following algorithm to estimate response times to traffic signal changes:

\begin{enumerate}
	\item Log the time when the next traffic signal in front of the driver changes from green to yellow.
	\item If the time headway between the driver and the traffic signal at the time of the signal change is large (e.g., over 10 seconds), stop looking for a reaction time.
	\item If the leading car is also before the light, check to see if it decelerates.  If it does, stop looking for a reaction time.
	\item Check to see if the car decelerates.  The difference between the time when the car decelerates and when the signal changed is the response time.
	\item Follow up to see if the car turns at the intersection.  If it does, ignore the measured reaction time.
\end{enumerate}
In order to be successful in tuning ITS algorithms to individual drivers, we will need a model which provides us with an estimate of the average driver's brake reaction time as well as the individual driver's response time.  The mix of drivers on the road is constantly changing, with new drivers joining and other, usually older, drivers leaving.  Thus when there is no information on an individual, the average response times can be used.  As more information about an individual driver's response times becomes available, the system can switch from the general estimate of brake response time to the individual driver's estimated brake response time.
\section{Estimating the Distribution of Potential Brake Response Times}
\label{estimation}

\subsection{General Discussion}

In this section we discuss the construction of a statistical model for the distribution of brake response times, and how this model can be used to estimate the distribution of potential brake response times for a particular individual.

As discussed earlier, previous researchers have consistently found that reaction times are skewed right and are approximated well by a lognormal distribution.  The plot of brake reaction times reproduced in Fig. \ref{fig:GohWongRTPlot} above confirms that for a fixed value of time headway, the distribution of brake reaction times across individuals is skewed right.  It is reasonable to assume that brake reaction times are skewed right within individuals as well.  We therefore adopt a lognormal model for brake reaction times, modelling the logarithm of the observed BRT as normally distributed conditional on the time headway.

This lognormal model also has the advantage of automatically correcting for some differences in the variance of the BRT distribution at different time headways and across individuals.  From the plot in Fig. \ref{fig:GohWongRTPlot}, we can see that as the time headway increases, the mean BRT and the variance of the BRTs both increase.  Similarly, it seems likely that some individuals have lower or higher mean reaction times than other drivers, and that the variance in the BRT distribution varies across individuals as well.  Specifically, it is likely that individuals with a low mean reaction time also have a low variance in their reaction times, whereas individuals with a high mean reaction time also have a high variance in their reaction times.  These differences in the variance of brake reaction times will be approximately corrected by modelling the logarithm of the BRT.

It also seems likely that the mean and variance of the brake response time distribution depend on several other variables.  An important factor that will be accounted for in our model is the stimulus type (e.g. traffic signal vs. lead car decelerates).  
Reaction times also depend on a large number of other factors such as weather conditions and demographic characteristics of the driver.  However, these variables will not generally be available to the accident warning system, so their effects will be absorbed into the error term of our model.


\subsection{The Model}

Using just the time headway as an explanatory variable, the general ideas above can be formalized in the following model:

\begin{align}
\nonumber \by_{d} &\sim N(X\beta + X\gamma_{d}, \sigma^2 I ) \nonumber \\
\gamma_{d} &\sim N(0, \Sigma_\gamma) \label{ModelStatement}
\end{align}

In this model,
\begin{itemize}
	\item $d$ indexes the driver
	\item $\by_{d}$ is a vector of the logarithms of observed reaction times for a particular driver.
	\item $X$ is a matrix of covariates, detailed further below.
	\item $\beta$ is a fixed vector of unknown coefficients.
	\item $\sigma^2$ is an unknown scalar.
	\item $\gamma_d$ is a random vector of unknown coefficients.
	\item $\Sigma_\gamma$ is an unknown matrix.
\end{itemize}

\begin{figure}[!t]
	\centering
\framebox{\parbox{2.5in}{\includegraphics[width=2.5in]{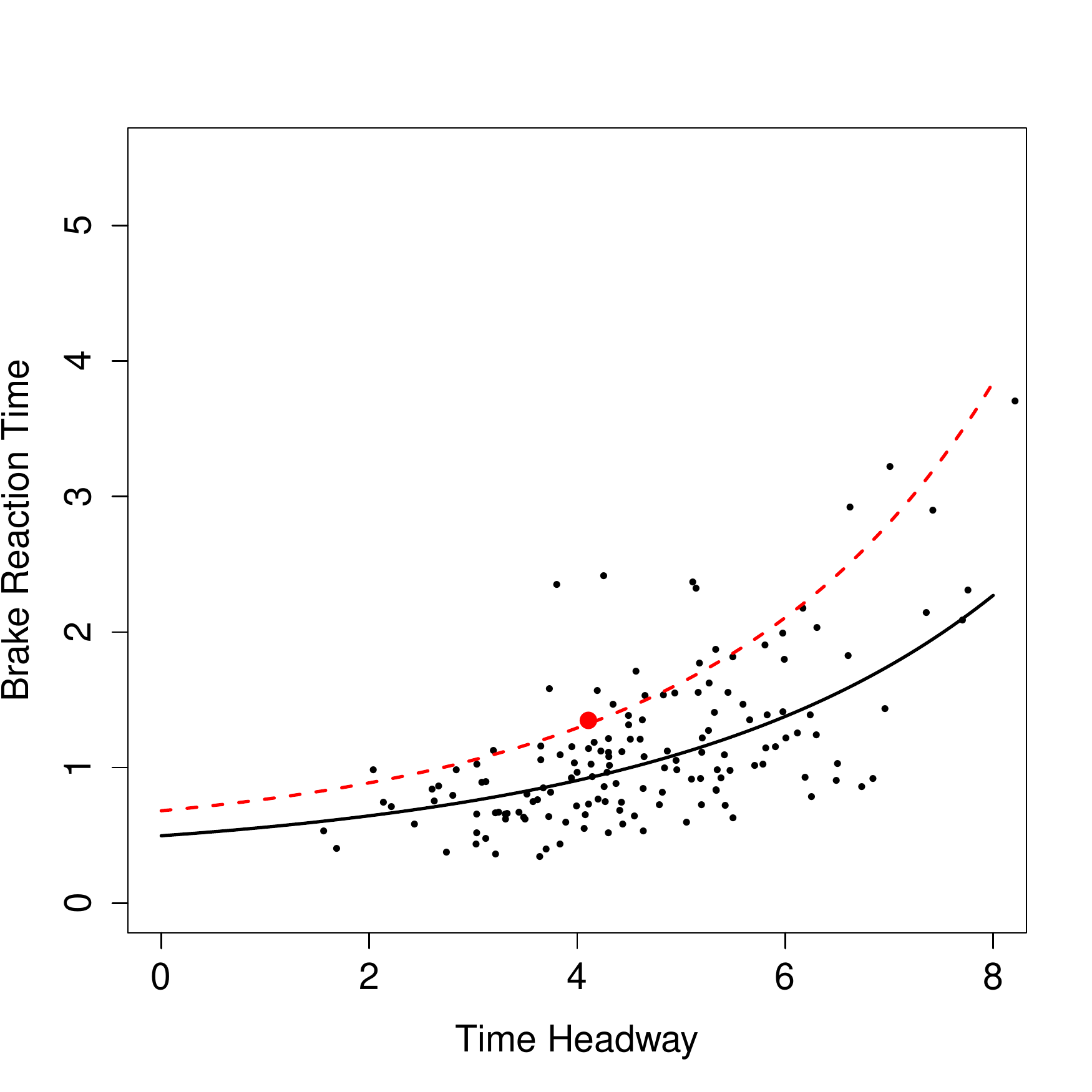}
}}
		\quad
\framebox{\parbox{2.5in}{\includegraphics[width=2.5in]{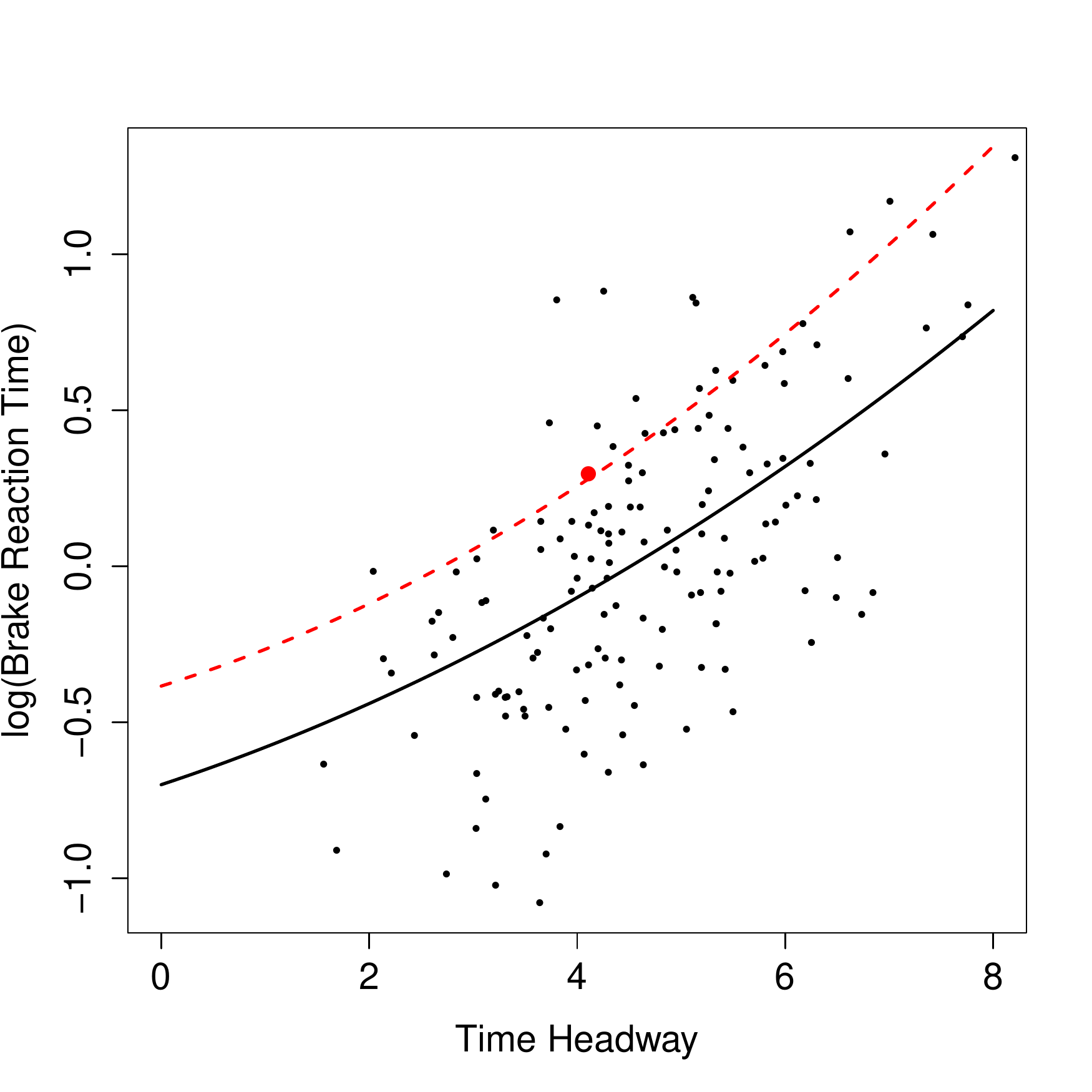}
}}
	\caption{An illustration of the model based on a simulated data set.  All parameters were chosen for the simulation so that the simulated data would be reasonably similar to that in Fig. \ref{fig:GohWongRTPlot}, from Goh and Wong \cite{GohWong:DriverPRTDuringSignalChange}.  Each plot shows simulated data for just one stimulus type.  The black curve represents the population-average relationship between time headway and brake reaction time, $X\beta$.  The red curve represents the relationship between time headway and brake reaction time for one individual, $X(\beta + \gamma)$.  The red point is an observation for that driver.}
	\label{fig:ModelIllustrationSim}
\end{figure}

The basic idea of this model is that, conditional on the time headway, the distribution of BRTs for an individual driver has a mean which is given by an overall population mean, $X\beta$, plus an offset due to the particular characteristics of that driver, $X\gamma_d$.  This is illustrated in Fig. \ref{fig:ModelIllustrationSim}.  It is assumed that the parameters $\gamma_d$ determining the individual's offset to the overall mean follow a multivariate Normal distribution in the population.  This is a linear mixed effects model; three recommended references with further information about these models are McCulloch et al., Ravishanker and Dey, and Searle et al \cite{McCullochetal:GLMM}, \cite{Searleetal:VC}, \cite{RavishankerDey:LMT}.
A key assumption made in this model specification is that after the log transformation, the covariance matrix $\cov[\by_{d}]$ has the simple form $\sigma^2 I$.  This assumption could fail to hold in a number of ways, but it makes the calculations much easier.

We now consider the form of the mean $X(\beta + \gamma_d)$ in more detail.  From the plot in Fig. \ref{fig:GohWongRTPlot}, we saw that the mean brake reaction time was an increasing function of time headway.  Since the logarithm is a monotonically increasing function, it follows that the logarithm of the BRT is also an increasing function of time headway.  For flexibility, we allow the possibility that the log BRTs are a quadratic function of time headway.  We also allow for the possibility that the relationship between time headway and BRT is slightly different for each of the different stimulus types.  For instance, it could be that drivers have a faster BRT at low time headways and the average BRT increases more rapidly as a function of time headway when the stimulus is a lead car braking than when it is a traffic signal changing to yellow.  These considerations lead to the following possible form of the mean log-BRT as a function of time headway:
\begin{equation}
\nonumber E[y_{dsi}] =
\end{equation}
\begin{equation} \label{quad}
\beta_{s,0} + \beta_{s,1} t_{dsi} + \beta_{s,2} t_{dsi}^2 + \gamma_{d,s,0} + \gamma_{d,s,1} t_{dsi} + \gamma_{d,s,2} t_{dsi}^2
\end{equation}
In equation (\ref{quad}), $d$ indexes the driver, $s$ indexes the stimulus type, and $i$ indexes the observation (so if we have 5 different BRT observations for a particular driver and stimulus type, $i$ will vary from 1 to 5).  As before, $y_{dsi}$ is the log brake reaction time, and $t_{dsi}$ is the time headway at the time of the stimulus.  The subscript $s$ on the $\beta$ and $\gamma$ terms indicate that the values of those coefficients depend upon the stimulus type $s$.  To make this concrete, if this mean function is adopted and there are $S = 3$ different stimulus types under consideration with $n_{ds}$ observations for driver $d$ under stimulus type $s$, $\beta$ and $\gamma_d$ are $9 \times 1$ vectors and the portion of the $X$ matrix corresponding to observations for driver $d$ will be of the following form:

$$\begin{bmatrix} 	\centering
1 & t_{d11} & t_{d11}^2 & 0 & 0 & 0 & 0 & 0 & 0 \\
 1 & t_{d12} & t_{d12}^2 & 0 & 0 & 0 & 0 & 0 & 0 \\
\vdots & \vdots & \vdots & \vdots & \vdots & \vdots & \vdots & \vdots & \vdots \\
 1 & t_{d1n_{d1}} & t_{d1n_{d1}}^2 & 0 & 0 & 0 & 0 & 0 & 0 \\
 0 & 0 & 0 & 1 & t_{d21} & t_{d21}^2 & 0 & 0 & 0 \\
\vdots & \vdots & \vdots & \vdots & \vdots & \vdots & \vdots & \vdots & \vdots \\
 0 & 0 & 0 & 1 & t_{d2n_{d2}} & t_{d2n_{d2}}^2 & 0 & 0 & 0 \\
 0 & 0 & 0 & 0 & 0 & 0 & 1 & t_{d31} & t_{d31}^2 \\
\vdots & \vdots & \vdots & \vdots & \vdots & \vdots & \vdots & \vdots & \vdots \\
 0 & 0 & 0 & 0 & 0 & 0 & 1 & t_{d3n_{d3}} & t_{d3n_{d3}}^2
\end{bmatrix}$$

\subsection{Training the Model: A Fit Using Data from Driving Simulations}

For training the model, we assume data are gathered for $D$ subjects in a driving simulation.  If possible, we prefer to gather data from real drivers on the road, but this is likely to be too difficult to be feasible.  This being the case, we will take precautions to address concerns about using results from a driving simulation to learn about response times for drivers in real life driving situations. The subjects in the study will be a representative sample of the overall population of drivers who will be using the accident warning system.  Brake responses for each subject will be elicited at a variety of levels of expectancy.  To improve the statistical analysis, responses will also be collected at a range of time headways for each stimulus type.  To separate the effects of expectancy and any other variables that may be included in the model, the combinations of these factors will be randomized (for example, we will have some observations where the braking stimulus was more and less surprising at different levels of the time headway variable).

For each driver, we have multiple observations of reaction times for each stimulus type.  These data can be used to estimate the unknown quantities $\beta$, $\sigma^2$, and $\Sigma_\gamma$ in this model using standard statistical techniques implemented in the lmer function of the lme4 library in R.  We will use a subscript of $(tr)$ to indicate quantities obtained from this training data set; in particular, let $X_{(tr)}$ be the covariate matrix obtained using data from this data set and denote the estimates by $\widehat{\beta}_{(tr)}$, $\widehat{\sigma}^2_{(tr)}$, and $\widehat{\Sigma}_{\gamma(tr)}$.  $\widehat{\beta}_{(tr)}$ can be written as $\widehat{\beta}_{(tr)} = (X_{(tr)}' V_{(tr)}^{-1} X_{(tr)} )^{-} X_{(tr)}' V_{(tr)}^{-1} \by_{(tr)}$, where $V_{(tr)} = \cov(\by_{(tr)}) = X_{(tr)} \Sigma_\gamma X_{(tr)}' + \sigma^2 I$ and the superscript $^{''-''}$ denotes a generalized inverse.  The estimates $\widehat{\sigma}^2_{(tr)}$ and $\widehat{\Sigma}_{\gamma(tr)}$ can be found through numerical maximum likelihood techniques.
A study conducted by McGehee et al. has found that the population average brake response time was about 0.3 seconds faster in driving simulations than it was in real life driving studies \cite{McGeheeetal:DriverRTSimvsReal}.  This difference was found at time headways of approximately 2 seconds.  It is difficult to account for this effect in a rigorous way, especially since this observed difference may be due in part to methodological differences between the simulator trials and the real car driving trials.  One ad hoc solution would be to increase the estimated value of $\widehat{\beta}_{0, (tr)}$ by an amount such that the estimated population mean reaction time at a time headway of 2 seconds increases by 0.3 seconds.

\subsection{Real Time Estimation of the PBRT Distribution for One Driver}
\label{subsubsec:FreqRealTimeReactionDistEst}

We estimate the distribution of PBRTs for a particular driver in two steps.  First, we establish the relationship between the covariates and BRT for that driver.  Then we use this relationship to estimate the distribution of PBRTs by using values of the covariates at which the BRT does not include an intentional delay to braking.

\subsubsection{Estimating the Relationship Between Time Headway and BRT for One Driver}

As data are gathered in real time for an individual driver $d^*$, our goal is to estimate the driver's offset $\gamma_{d^*}$ to the population-average regression coefficients $\beta$.  This is estimated by the Best Linear Unbiased Predictor (BLUP).

Intuitively, we might expect that if a particular driver has a higher than average brake response time in one stimulus type, they are likely to have a higher than average brake response time in other stimulus types as well.  Similarly, if they are particularly sensitive to the time headway in one situation, they are more likely to be sensitive to the time headway with other stimulus types.  This intuition suggests that the covariance matrix $\Sigma_\gamma$ will have non-zero off-diagonal entries; that is, there is some degree of correlation among the $\gamma_d$ coefficients.  Because of this correlation, observations from one stimulus type can give us information about the coefficients in the other stimulus types.  For example, if we make some observations of driver brake response times in the traffic light setting which give positive estimates of the $\gamma_d$ coefficients for that stimulus, a positive correlation between the coefficients might lead to positive estimates of the coefficients for other stimuli as well.

To reduce the computational complexity of computing the BLUP, we assume that the information about the unknowns $\beta$, $\sigma^2$, and $\Sigma_\gamma$ that is provided by the training data set from the driving simulator is much greater than the information provided by the data from this individual driver.  That is, the estimates $\widehat{\beta}_{(tr)}$, $\widehat{\sigma}^2_{(tr)}$, and $\widehat{\Sigma}_{\gamma(tr)}$ obtained from the training data set above are very similar to what we would obtain if we estimated them using the combined training data set with the observations for this driver.  If this assumption holds, we can approximate the BLUP using the estimates of these quantities found with the training data set, which saves the computational effort of re-fitting the model every time we observe a new reaction time.

Let $X_{d^*}$ be the covariate matrix $X$ as in the full model, but formed using only the data from driver $d^*$.  The BLUP of $\gamma_{d^*}$ is
\begin{equation}
\nonumber \tilde{\gamma}_{d^*} = \Sigma_\gamma X_{d^*}' V_{d^*}^{-1} (\by_{d^*} - X_{d^*} \widehat{\beta})
\end{equation}
where $V_{d^*} = \cov(\by_{d^*}) = X_{d^*} \Sigma_\gamma X_{d^*}' + \sigma^2 I$.  Ordinarily $\widehat{\beta}$ would be estimated from all of the data, but by our assumption above we will instead use the estimate $\widehat{\beta}_{(tr)}$.  The formula for the BLUP still involves the unknowns $\sigma^2$ and $\Sigma_\gamma$.  We estimate the BLUP by plugging in the estimates of these quantities obtained from the training data above.  Denoting this estimated BLUP by $\hat{\gamma}_{d^*}$, we have:
\begin{equation}
\nonumber \hat{\gamma}_{d^*} = \widehat{\Sigma}_{\gamma(tr)} X_{d^*}' \widehat{V}^{-1}_{d^*} (\by_{d^*} - X_{d^*} \widehat{\beta}_{(tr)}),
 \end{equation}
 where
$ \widehat{V}_{d^*} = X_{d^*} \widehat{\Sigma}_{\gamma(tr)} X_{d^*}' + \widehat{\sigma}^2_{(tr)} I.$ The covariance matrix of the BLUP $\tilde{\gamma}_{d^*}$ is given by

\begin{equation}
\nonumber \cov(\tilde{\gamma}_{d^*}) = \cov(\Sigma_\gamma X_{d^*}' V_{d^*}^{-1} (\by_{d^*} - X_{d^*} \widehat{\beta}))
\end{equation}
\begin{equation}
\nonumber = \Sigma_\gamma X_{d^*}' V_{d^*}^{-1} (V_{d^*} - X_{d^*} \cov(\widehat{\beta}_{(tr)}) X_{d^*}') V_{d^*}^{-1} X_{d^*} \Sigma_\gamma
\end{equation}
To estimate the covariance matrix of $\hat{\gamma}_{d^*}$, we plug our approximation to $\widehat{\beta}$, $\widehat{\beta}_{(tr)}$, and our estimates of $\sigma^2$, $\Sigma_\gamma$, and $\cov(\widehat{\beta}_{(tr)})$ into this formula.  Denote this estimated covariance matrix by $\widehat{\Sigma}_{\hat{\gamma}_{d^*}}$.

When no data have been gathered yet, the best predictor is just the vector 0, with covariance matrix $\Sigma_\gamma$. In this case, the estimated mean for the individual is equal to the estimated mean for the population of all drivers.

\subsubsection{Obtaining the Estimated PRBT Distribution}

\begin{figure}[!t]
	\centering
\framebox{\parbox{2.5in}{\includegraphics[width=2.5in]{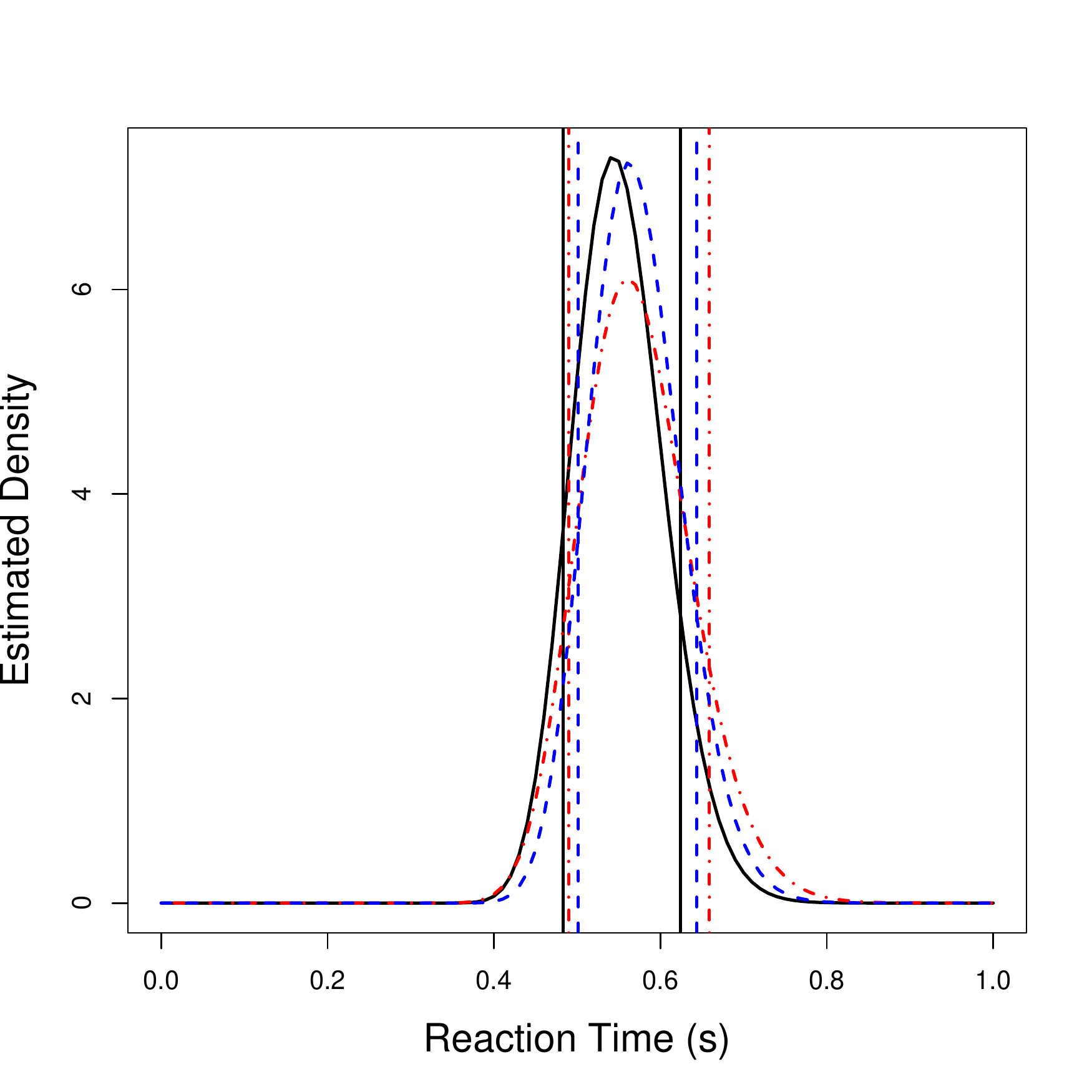}
}}
\caption{Estimates of the distribution of PBRTs for an individual obtained in a simulation.  The black curve represents the individual's ``true'' response time distribution.  The blue curve is the estimated distribution when the variance is taken to be $\hat{\sigma}^2$.  The red curve is the estimated distribution when the variance estimate includes a term for uncertainty in $\widehat{\beta}$ and $\widehat{\gamma_{d^*}}$.  The vertical lines are at the $10^{th}$ and $90^{th}$ percentiles.}
\label{fig:SimRTDistEst}
\end{figure}

The final step is to estimate the distribution of potential brake response times for an individual driver, not including any delays.  For the suggested model form above using a quadratic function of time headway, the intuitive idea is to pick a specific time headway value $t^*$ at which the driver does not have enough time to delay braking, and use that time headway value to evaluate the mean function.  Based on the plots in Fig. \ref{fig:GohWongRTPlot}, it appears that $t^* = 1.5$ might be an appropriate value.  We can then estimate the mean of the driver's log-RTs by plugging $t^* = 1.5$ into the estimated mean function:
$\hat{\mu} = \hat{\beta}_{0} + \hat{\gamma}_{d^*, 0} + t^* (\hat{\beta}_{1} + \hat{\gamma}_{d^*, 1}) + (t^*)^2 (\hat{\beta}_{2} + \hat{\gamma}_{d^*, 2})$.  This provides an estimated mean for the log-reaction time.
There are several options for estimating the variance of the log-PBRT distribution. One simple idea would be to use the estimate $\widehat{\sigma}^2_{(tr)}$ of the quantity $\sigma^2$ in the model statement \ref{ModelStatement}.  However, this does not take into account the uncertainty in our estimate $\hat{\mu}$.  This uncertainty is captured by the prediction error, $(\widehat{\beta}_{(tr)} + \hat{\gamma}_{d^*}) - (\beta + \gamma_{d^*})$.  It can be shown that $\cov((\widehat{\beta}_{(tr)} + \hat{\gamma}_{d^*}) - (\beta + \gamma_{d^*})) = \cov(\widehat{\beta}_{(tr)}) + \cov(\hat{\gamma}_{d^*} - \gamma_{d^*}) - \cov(\widehat{\beta}_{(tr)}, \gamma_{d^*}') - \cov(\gamma_{d^*}, \widehat{\beta}_{(tr)})$, where
\begin{align}
\nonumber\cov(\hat{\gamma}_{d^*} - \gamma_{d^*}) &= \Sigma_{\gamma} - \cov(\hat{\gamma}_{d^*})\\
\nonumber \cov(\hat{\gamma}_{d^*}) &=\\
\nonumber\Sigma_{\gamma} X_{d^*}' ( V^{-1}_{d^*} - V^{-1}_{d^*} X_{d^*} \cov(\widehat{\beta}_{(tr)})&X_{d^*}' V^{-1}_{d^*} )X_{d^*} \Sigma_{\gamma}  
\end{align}
\begin{equation}
\nonumber \cov(\widehat{\beta}_{(tr)}, \gamma_{d^*}') = \cov(\widehat{\beta}_{(tr)}) X_{d^*}' V^{-1}_{d^*} X_{d^*} \Sigma_{\gamma}
\end{equation}
This covariance can be estimated by plugging in estimates of the unknown quantities $V_{d^*}$, $\cov(\widehat{\beta}_{(tr)})$, and $\Sigma_{\gamma}$.  An estimate of the variance of the distribution of log-PBRTs which takes into account our uncertainty about the value of the mean is then
\begin{align}
\nonumber \begin{bmatrix} 1 & t^* & t^{*2} \end{bmatrix} \widehat{\cov}((\widehat{\beta}_{(tr)} + \hat{\gamma}_{d^*}) &- (\beta + \gamma_{d^*}))\begin{bmatrix} 1 & t^* & t^{*2} \end{bmatrix}'\\
\nonumber &+ \hat{\sigma}^2_{(tr)}
\end{align}
When we do not yet have any data, the adjusted variance estimate is
\begin{equation}
\nonumber \begin{bmatrix} 1 & t^* & t^{*2} \end{bmatrix} \widehat{\Sigma}_\gamma \begin{bmatrix} 1 & t^* & t^{*2} \end{bmatrix}'+ \hat{\sigma}^2_{(tr)}.
\end{equation}
The plot in Fig. \ref{fig:SimRTDistEst} shows the resulting distribution estimates obtained in a simulation when these variance estimates are used as the parameters of the distribution of PBRTs.  From this plot we can see that the estimates taking into account uncertainty in the coefficient estimates are more conservative.  On the scale of these simulation results, the difference in the percentiles obtained from these estimates is just a fraction of a second, but the difference could be more significant with real data.  We will use the more conservative value for the estimated variance since it more accurately reflects what we know about the distribution of response times based on the available data.

\begin{figure}[!t]
	\centering
\framebox{\parbox{2.5in}{\includegraphics[width=2.5in]{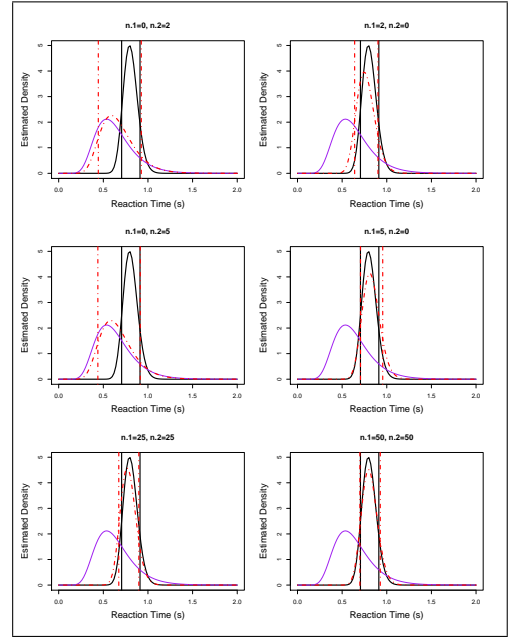}
}}
\caption{Estimates of the distribution of PBRTs for an individual obtained in a simulation with different sample sizes.  The black curve represents the individual's ``true'' response time distribution.  The purple curve represents the distribution of reaction times in the population, which is used as an estimate when the sample size is 0.  The red curve is the estimated distribution.  The vertical lines are at the $10^{th}$ and $90^{th}$ percentiles.}
\label{fig:SimRTDistEstvsSampleSize}
\end{figure}

Fig. \ref{fig:SimRTDistEstvsSampleSize} shows how the estimated reaction time distribution changes with the sample size and the allocation of the sample among the different stimulus types.  These results are dependent upon the parameter values used in the simulation, but they illustrate that observed reaction times for the stimulus type that is used in estimating the PBRT distribution contribute more information than observations in other stimulus types.  This will generally be the case, but our simulation likely shows an extreme example since the correlation among the gamma coefficients for different stimulus types is very low in the simulation.  It could be helpful to run a simulation like this once the training data has been gathered to determine what sample sizes are necessary to get good estimates of the ``true'' PBRT distribution.

We note that computation of the estimated PBRT distribution requires only the operations of matrix inversion and matrix multiplication.  The matrix which must be inverted is $\widehat{V}_{d^*}$, which has dimension $n_{d^*}$, the number of observations for driver $d^*$.  The inversion operation has computational complexity $O(n_{d^*}^3)$.  All of the matrix multiplication operations are between matrices of dimension $9 \times 1$, $9 \times 9$, $9 \times n_{d^*}$, $n_{d^*} \times 1$, or $n_{d^*} \times 1$.  Because multiplying an $n \times m$ matrix by an $m \times k$ matrix has complexity $O(nmk)$, this implies that the complexity of the ``worst'' matrix multiplication operation is $O(9n_{d^*}^2)$ (for the product $X_{d^*}' \widehat{V}_{d^*}^{-1}$).  Therefore the whole computation has complexity $O(n_{d^*}^3)$ when $n_{d^*} > 9$.

\subsection{Estimated PRBT Distribution vs Population Distribution}
\label{subsec:FreqRealTimeReactionDistEst}
\begin{figure}[!t]
	\centering
\framebox{\parbox{3in}{\includegraphics[width=3in]{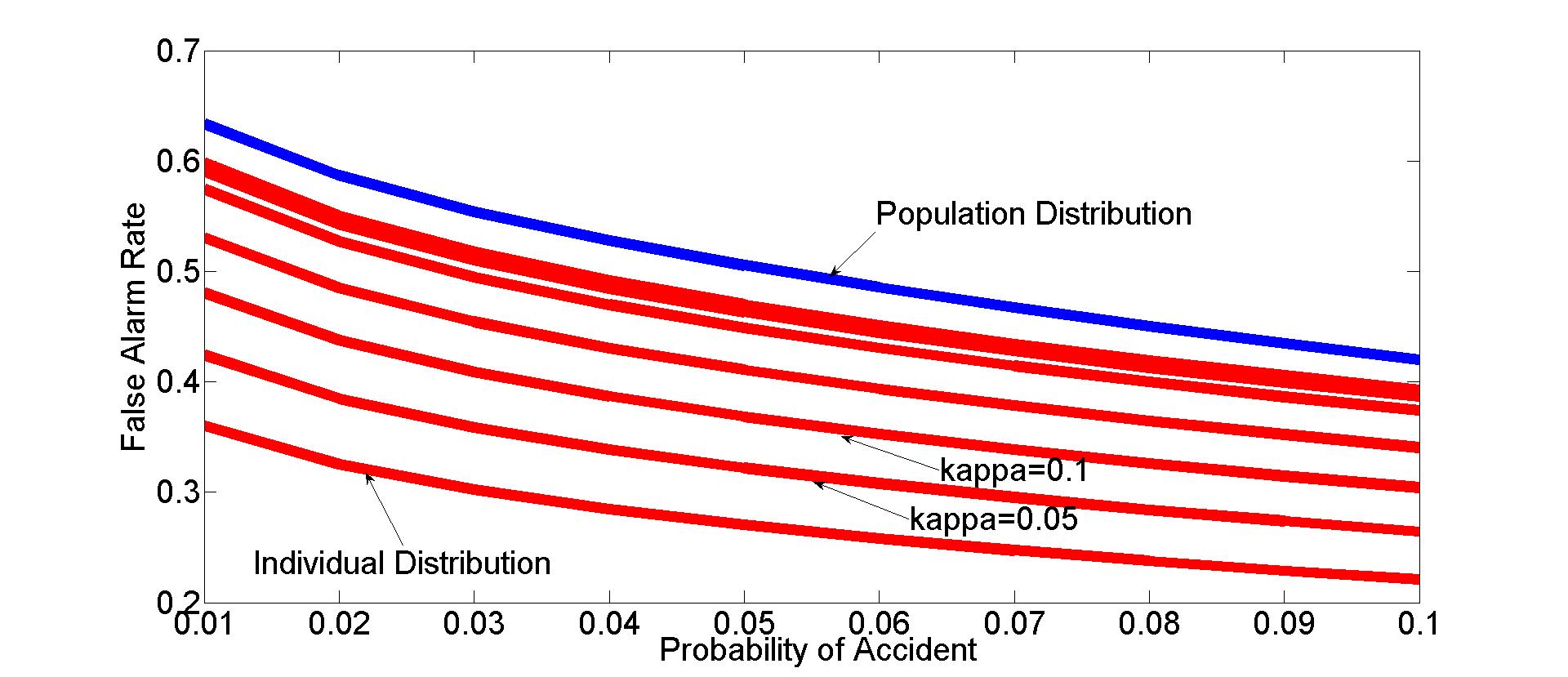}
}}
\caption{A conventional system versus a customized system. Figure shows the false alarm rate (y axis) versus the probability of accident (x axis), the percentage of possible collisions that the system fails to give warning about, using population and individual PBRT distributions. Population distribution $= lnN(0.17, 0.44^2)$, based on results from \cite{Koppa:HumanFactors},
and individual PBRT distribution based on results of \cite{GohWong:DriverPRTDuringSignalChange}.}
\label{fig:FAR_POA}
\end{figure}

Our goal in this section is to make a comparison between two types of systems.
\begin{enumerate}
\item
Conventional systems which use the population distribution.
\item
Customized systems which use an individual driver's distribution.
\end{enumerate}
We described our method \cite{Rakhshan} for estimating the PBRT distribution which can then be entered into collision warning algorithms. Next, we want to relate this distribution to the distribution of PRTs for the population to show how collision warning algorithms benefit from taking the estimated distribution for an individual driver into account. As discussed earlier, previous researchers have consistently found that reaction times are skewed right and are approximated well by a lognormal distribution.
It is reasonable to assume that brake reaction times are skewed right within individuals as well.
\begin{equation}\label{eq:trunc}
 f{(x|\mu)}=\frac{1}{ x\sigma\sqrt{2 \pi}} e^{-\frac{(\ln{x}-\mu)^2}{2\sigma^2}}
 \qquad  \textrm{for} \qquad  x\geq0
\end{equation}
In equation \ref{eq:trunc}, we let $X$ be a random variable representing the PRT
for an individual driver. Also, to analyze the situation, we let the mean of the distribution for an individual driver, denoted by $\mu$ , be a random variable. This assumption will represent the fact
that different drivers in the population have different PRT means, and that more drivers
have means within certain ranges than others.
For our model of driver PRTs to be reasonable, the marginal distribution $f_X(x)$ should closely match the log-normal distribution for the overall population found in the literature.
Once a distribution for a driver's reaction time has been established, we would like to use
this information to improve safety and minimize the rate of false alarms. One simple method
for doing this would be to give the driver a warning when they are approaching an obstacle,
and there will not be enough time for them to react otherwise. As we mentioned, \cite{Koppa:HumanFactors} established that the distribution of PRTs of drivers
reacting to surprise events follows a log-normal curve with parameters $\mu= 0.17$ and $\sigma = 0.44$. Since failing to give a warning when one is needed could be very dangerous, we will assume
that the percentage of possible collisions that the system fails to provide warning for is fixed at a small number (e.g. at 1\%), and then try to minimize the frequency of false alarms that the system gives subject to this constraint.
If the system detects that the driver has less than his or her PRT to react to an obstacle, it should give the driver a warning.
 We can only state the probability that any PRT is above or below a certain value. Thus, the constraint states that we must calculate some threshold $T_t$ above which there is only $1\%$ chance that a PRT will be , and send a warning whenever a driver has less than this amount of time to react.
Therefore, we can calculate the threshold to send the warnings using  the distribution for the entire population:
\begin{equation}
\nonumber P(X \leq T_t)=\Phi\left(\frac{\ln(T_t)-0.17}{0.44}\right)=1- \text{prob. of accident}\\
\end{equation}
\begin{equation}
\nonumber \text{If probability of accident=1\%}
\quad\Rightarrow T_t=e^{1.9} \approx 3.3
\end{equation}
Also, we can calculate warning threshold using the distribution for an individual driver:
\begin{equation}
\nonumber P(X \leq T_t)=\Phi\left(\frac{\ln(T_t)-\mu}{\sigma}\right)=1- \text{prob. of accident}\\
\end{equation}
Now that we have established the thresholds for sending collision warnings, we can calculate
the false alarm rates that will result from using the different systems. A false alarm occurs
whenever a warning is sent, but it is not needed. To best explain this problem, let us consider the scenario that
 a vehicle is following another vehicle on a one-lane roadway when the lead
 vehicle suddenly begins to decelerate to avoid an unexpected obstacle.
Suppose that the system has
calculated that the following driver has $t$ seconds to react, and that $t$ is less than $T_t$, therefore a warning
has been sent. Then, the false alarm rate is the probability that the driver's reaction time,
X, will be less than $t$. Let $F_X(x)$ denote the cumulative distribution function for this distribution then $F_X(t)$ is the total false alarm rate.

\begin{equation}\label{eq:FAR_total}
\nonumber P(X \leq T) \:=\int_{0}^{T_t} \int_{0}^{t} \frac{1}{ x\sigma\sqrt{2 \pi}} e^{-\frac{(\ln{x}-\mu)^2}{2\sigma^2}}\frac{1}{T_t} dxdt\\
\end{equation}
 It is clear from Fig. \ref{fig:FAR_POA} that when we use the population brake reaction the false alarm rate is higher by almost a factor of two than when we use the individual driver's brake reaction time. 
  Therefore, safety applications could potentially take full advantage of being customized to an individual's characteristics. Regardless, there is an observable tradeoff between the false alarm rate and the probability of an accident, one that cannot be remediated by obtaining estimates of an individual's brake reaction time. 
It's worth mentioning that false alarms are not evenly distributed across the population. Drivers with fast reaction times will have very high false alarm rates, but drivers with slow reaction times will have lower rates. 
  If the warnings turn out to be false alarms too frequently, drivers may begin to ignore them, and since novice drivers receive the most false alarms, the danger to ignore the safety system is higher for them. While it is known that collision warnings can offer a great help to older drivers, our method benefits novice drivers as well. If it were simply the case that novice drivers were careless, warnings might be of little use.  But, the existing research suggests that many novice drivers are clueless, not careless, e.g. in \cite{Pollatsek}. Thus, it is of vital importance to minimize false alarms so that the system sends warnings only when 
 it is needed. 

 Also, we need to take into account the estimation errors since in real life our estimations are not accurate.
Fig. \ref{fig:FAR_POA} shows that by increasing the error (equation \ref{eq:Error}) of the estimated distribution, the individual curve approaches the population curve. Our error model can be described as:
 \begin{align}\label{eq:Error}
\nonumber \hat{X}&=X+e\\
e &\sim N(0,\kappa^2I)
\end{align}
 High level of errors in estimating the distribution of individual drivers is almost equivalent to the case when no individual sample is available; thus, the system uses the population distribution similar to the conventional systems.

\section{Conclusion And Future Work} \label{conc}
In this paper we discussed the need to adapt collision warning systems
to drivers' individual characteristics and proposed a method for doing this customization
by estimating the distribution of potential brake response times for an
individual driver in real time. Collision warning systems generally rely solely on the distribution of the entire population of drivers, thereby ignoring the distinct characteristics of individual drivers. They may frustrate the drivers with the overly high numbers of false alarms, causing them to ignore warnings or even disable the system. If drivers are distracted by overly frequent warnings, the safety benefits of the system are compromised or even lost.

Our proposed method uses a statistical model that was developed based on previously published results about the
population-level brake response times.  This model has not yet
been validated with data that includes multiple reaction times for each
driver. However, we demonstrated why employing this method will result in reducing the rate of traffic collisions, thereby dramatically improving the safety benefits for all drivers. In our future work, we will collect this data, fine-tune the model, and apply it to a collision warning system.
\begin{IEEEbiographynophoto}{Ali Rakhshan}
is currently pursuing
a Ph.D. degree in electrical and computer engineering at the University of Massachusetts, Amherst. With a broader scope
of communication systems, his main field of interest is Vehicular Ad Hoc Networks (VANETs).
\end{IEEEbiographynophoto}
%
\begin{IEEEbiographynophoto}{Evan Ray}
is a Ph.D. student in the Statistics program at the University of Massachusetts, Amherst.  He previously earned a M.S. in Statistics from the University of Massachusetts, Amherst and a B.S. in Mathematics from the University of Massachusetts, Boston.  His research interests include state-space models and measurement error.
\end{IEEEbiographynophoto}
\begin{IEEEbiographynophoto}{Hossein Pishro-Nik}
is an Associate
Professor of electrical and computer engineering
at the University of Massachusetts, Amherst. He
received a B.S. degree from Sharif University of
Technology, and M.Sc. and Ph.D. degrees from the
Georgia Institute of Technology, all in electrical
and computer engineering. His research interests
include the mathematical analysis of communication
systems, in particular, error control coding,
wireless networks, and vehicular ad hoc networks.
His awards include an NSF Faculty Early Career
Development (CAREER) Award, an Outstanding Junior Faculty Award from
UMass, and an Outstanding Graduate Research Award from the Georgia
Institute of Technology.
\end{IEEEbiographynophoto}
%
%
%
%
%
%
%
%

\begin{thebibliography}{1}

\bibitem{USDoT:TrafficFatalities}
 United States Department of Transportation.  National Highway Traffic Safety Administration \emph{Early Estimate of Motor Vehicle Traffic Fatalities in 2011.}\hskip 1em plus
  0.5em minus 0.4em\relax : DOT HS 811 604.  Washington, DC, 2012.

\bibitem{Koppa:HumanFactors}
R.J.~Koppa, \emph{Human Factors}, In Nathan H. Gartner, Carroll J. Messer, and Ajay K. Rathi (Eds.)., Revised Monograph on Traffic Flow Theory (Ch.3).,.\hskip 1em plus 0.5em minus 0.4em\relax : <http://www.fhwa.dot.gov/publications/, 2005.

\bibitem{Green:HowLongToStop}
M.~Green, \emph{How long does it take to stop?}, Methodological analysis of driver perception-brake times, ,.\hskip 1em plus 0.5em minus 0.4em\relax : Transportation Human Factors,2(3), 195-216, 2000.

\bibitem{GohWong:DriverPRTDuringSignalChange}
P.~Goh, and Y.~D.Wong, \emph{Driver perception response time during the signal change interval}\hskip 1em plus
  0.5em minus 0.4em\relax : Appl Health Econ Health Policy, 2004.

\bibitem{Rakha:DriBeh-conf}
J.R.~ Setti, H.~Rakha, and I.~ El-Shawarby \emph{Analysis of Brake Perception-Reaction Times on High-Speed Signalized Intersection Approaches}\hskip 1em plus
  0.5em minus 0.4em\relax : Intelligent Transportation Systems Conference, ITSC '06. IEEE, 2006.

 \bibitem{MaxwellWood:ReviewTrafficSignal}
A.~Maxwell, and K.~Wood, \emph{Review of Traffic Signals on High Speed Road. Paper presented at the Europian Transport Conference},\hskip 1em plus 0.5em minus 0.4em\relax : http://www.etcproceedings.org/paper/review-of-traffic-signals-on-high-speed-roads, Accessed 12/2/12.

\bibitem{WortmanMatthias:EvalDriverBehavior}
R.H.~Wortman, and J.S.~ Matthias, \emph{An Evaluation of Driver Behavior at Signalized Intersections},\hskip 1em plus 0.5em minus 0.4em\relax : Phoenix: Arizona Department of Transportation, 1983.

\bibitem{ZhangBham:EstDriverRT}
X.~Zhang, and G.H.~ Bham, \emph{Estimation of driver reaction time from detailed vehicle trajectory data.},\hskip 1em plus 0.5em minus 0.4em\relax : Proceedings of the 18th IASTED International Conference: modeling and simulation,574-579, 2007.

\bibitem{Rakha:DriBeh}
H.~Rakha, I.~ El-Shawarby, and J.R.~ Setti, \emph{Characterizing Driver Behavior on Signalized Intersection Approaches at the Onset of a Yellow-Phase Trigger.},\hskip 1em plus 0.5em minus 0.4em\relax : IEEE Transactions on Intelligent Transportation Systems,Volume:8 ,  Issue: 4, 630 - 640, 2007.

\bibitem{MaAndreasson:ReactionDelayEst}
X.~Ma, and Ingmar~ Andr\'easson, \emph{Driver reaction delay estimation from real data and its application in gm-type model evaluation},\hskip 1em plus 0.5em minus 0.4em\relax : Transportation Research Record: Journal of the Transportation Research Board, 1965, 130 - 141, 2006.

\bibitem{Ahmed:ModelingDrivers}
K.I.~Ahmed, \emph{Modeling drivers' acceleration and lane changing behavior},\hskip 1em plus 0.5em minus 0.4em\relax : Ph.D. Dissertation, Massachusetts Institute of Technology, 2006.

\bibitem{McCullochetal:GLMM}
C.E.~McCulloch, S.R.~Searle, and J.M.~Neuhaus \emph{Generalized, Linear, and Mixed Models},\hskip 1em plus 0.5em minus 0.4em\relax : 2nd ed.  Hoboken: Wiley, 2008.

\bibitem{Searleetal:VC}
S.R.~Searle, G.~Casella, and C.E.~McCulloch\emph{Variance Components},\hskip 1em plus 0.5em minus 0.4em\relax: New York: Wiley, 1992.

\bibitem{RavishankerDey:LMT}
N.~Ravishanker, and D.K.~Dey, \emph{A First Course in Linear Model Theory},\hskip 1em plus 0.5em minus 0.4em\relax : Boca Raton: Chapman \& Hall/CRC, 2002.

\bibitem{McGeheeetal:DriverRTSimvsReal}
D.V.~McGehee, E.N.~Mazzae, and G.H.S.~Baldwin, \emph{Driver Reaction Time in Crash Avoidance Research: Validation of a Driving Simulator Study on a Test Track.},\hskip 1em plus 0.5em minus 0.4em\relax Proceedings of the International Ergonomics Association 2000 Conference,
320-323., 2000.

	
\bibitem{Pollatsek}
A.~Pollatsek, D.~Fisher, and A.K.~Pradhan, \emph{Identifying and Remediating Failures of Selective Attention in Younger Drivers},\hskip 1em plus 0.5em minus 0.4em\relax Current Directions in Psychological Science, 15,255-259, 2006.

\bibitem{Rakhshan}
A.~Rakhshan, H.~Pishro-Nik, and E.~Ray, \emph{Real-Time Estimation of the Distribution of Brake Response Times for
an Individual Driver Using Vehicular Ad Hoc Network},\hskip 1em plus 0.5em minus 0.4em\relax IEEE Intelligent Vehicles Symposium, 2014.

%
%
%
%
%
%
\bibitem{Globecom}
A.~Rakhshan, H.~Pishro-Nik, M.~ Nekoui and D.~Fisher, \emph{Tuning Collision Warning Algorithms to Individual Drivers for Design of Active Safety Systems}, Globecom 2013 Workshop – Vehicular Network Evolution (GC13 WS – VNE), 1342-1346, 2013.

\end{thebibliography}
\end{document}